\documentclass[
reprint,
superscriptaddress,
amsmath,
amssymb,
aps,
prl,
floatfix
]{revtex4-2}

\usepackage{graphicx}% Include figure files
\usepackage{dcolumn}% Align table columns on decimal point
\usepackage{bm}% bold math
\usepackage[colorlinks,citecolor=magenta]{hyperref}

\begin{document}

\title{Exceptional Points of Temporal Coherent Perfect Absorption in the Momentum Bandgap of a Photonic Time Crystal}

\author{Yuhang Li}
\affiliation{Beijing National Laboratory for Condensed Matter Physics, Institute of Physics, Chinese Academy of Sciences, Beijing 100190, China}
\affiliation{School of Physical Sciences, University of Chinese Academy of Sciences, Beijing 100049, China}
\author{Zilong Bao}
\affiliation{Beijing National Laboratory for Condensed Matter Physics, Institute of Physics, Chinese Academy of Sciences, Beijing 100190, China}
\affiliation{School of Physical Sciences, University of Chinese Academy of Sciences, Beijing 100049, China}
\author{Yu Zhuang}
\affiliation{Beijing National Laboratory for Condensed Matter Physics, Institute of Physics, Chinese Academy of Sciences, Beijing 100190, China}
\affiliation{School of Physical Sciences, University of Chinese Academy of Sciences, Beijing 100049, China}
\author{Jingwen Cui}
\affiliation{Beijing National Laboratory for Condensed Matter Physics, Institute of Physics, Chinese Academy of Sciences, Beijing 100190, China}
\affiliation{School of Physical Sciences, University of Chinese Academy of Sciences, Beijing 100049, China}
\author{Junda Wang}
\affiliation{Beijing National Laboratory for Condensed Matter Physics, Institute of Physics, Chinese Academy of Sciences, Beijing 100190, China}
\affiliation{School of Advanced Interdisciplinary Sciences, University of Chinese Academy of Sciences, Beijing 100049, China}
\author{Zhanchun Zuo}
\affiliation{Beijing National Laboratory for Condensed Matter Physics, Institute of Physics, Chinese Academy of Sciences, Beijing 100190, China}
\affiliation{School of Physical Sciences, University of Chinese Academy of Sciences, Beijing 100049, China}
\author{Xiulai Xu}
\email{xlxu@pku.edu.cn}
\affiliation{State Key Laboratory for Mesoscopic Physics and Frontiers Science Center for Nano-optoelectronics, School of Physics, Peking University, 100871 Beijing, China}
\affiliation{Peking University Yangtze Delta Institute of Optoelectronics, Nantong, Jiangsu 226010, China}
\affiliation{Collaborative Innovation Center of Extreme Optics, Shanxi University, Taiyuan, Shanxi 030006, China}
\author{Chenjiang Qian}
\email{chenjiang.qian@iphy.ac.cn}
\affiliation{Beijing National Laboratory for Condensed Matter Physics, Institute of Physics, Chinese Academy of Sciences, Beijing 100190, China}
\affiliation{School of Physical Sciences, University of Chinese Academy of Sciences, Beijing 100049, China}
%\date{\today}

\begin{abstract}

We report exceptional points (EPs) of temporal coherent perfect absorption (CPA) in the momentum bandgap of a photonic time crystal.
Confinement of the photonic time crystal in a cavity with finite length gives rise to discrete momentum modes, and therefore enables coherent matching of a single light source to a selected decaying Floquet mode.
Meanwhile, excitation of the corresponding growing Floquet mode is suppressed by destructive self-interference.
Two CPA branches coalesce at second-order EPs, leading to a quartic absorption lineshape and enhanced robustness against momentum detuning.
Phase control further reveals the nontrivial topology of the CPA manifold, in which CPA at the wavevector of maximum Floquet rate exhibits a double phase winding around a third-order EP.
Compared to conventional CPA EPs that rely on material absorption, temporal CPA EPs transfer the source energy to the external modulation, providing the basis for high energy efficiency and suppressed local heating.
These results indicate the great potential of photonic time crystals for coherent control of light in integrated nanophotonic devices.

\end{abstract}

\maketitle

% \section{Introduction}
Photonic time crystals are materials having a permittivity that periodically varies in time \cite{10.1364/AOP.525163}.
The temporal modulation couples various frequency components at the same wavevector.
Resonant coupling between these Floquet components results in parity-time symmetry breaking and opens a gap in momentum space \cite{10.1364/OPTICA.5.001390, 10.1126/sciadv.abo6220}.
Modes within the bandgap have complex Floquet frequencies and exhibit exponential growth or decay in time \cite{10.1038/s41566-024-01563-3, 10.1103/PhysRevLett.134.063801}.
These unique properties establish the photonic time crystal as an ideal platform for time-domain photonics, such as enhanced spontaneous emission \cite{10.1126/science.abo3324, 10.1103/5v2w-yg7v}, lasing \cite{10.1103/6gn2-2v9b, 10.1103/hh9h-qzpk}, dynamical Casimir effect \cite{10.1126/science.abo3324, 10.1038/s42005-019-0183-z}, temporal topological states \cite{10.1038/s41467-025-66154-4, 10.1103/PhysRevB.111.125421, 10.1021/acsphotonics.4c01785,10.1002/lpor.202100469,10.1038/s41467-025-56021-7}, and nonlinear wave dynamics \cite{10.1103/PhysRevLett.130.233801, 10.1103/q226-dkk4, 10.1038/s41377-025-01788-z, 10.1103/t58q-lmng}.

Beyond amplification, photonic time crystals also provide new degrees of freedom for coherent control and energy transfer in light–matter interactions.
Galiffi et al. have recently reported the selective excitation of decaying modes by controlling the relative phase of two source pulses  \cite{10.1038/s41566-025-01833-8}, extending the concept of coherent perfect absorption (CPA) \cite{10.1103/PhysRevLett.105.053901, 10.1126/science.1200735} to the temporal domain.
Spatial confinement of the photonic time crystal further enables resonant control of the interaction between discrete cavity modes and the temporal modulation \cite{10.1002/lpor.71318, 10.1103/m5ww-n7ws, 10.1038/s41377-026-02380-9, 10.1103/175n-lksd}.
The interplay between these spatial and temporal degrees of freedom paves the way for exploring non-Hermitian phenomena, such as exceptional points (EPs) and their associated topology \cite{10.1126/science.aar7709, 10.1038/s42254-022-00516-5}, and provides the basis for tunable and robust optical absorption \cite{10.1038/s41467-025-61282-3}.

Here we report temporal CPA EPs in the momentum bandgap of a photonic time crystal by combining temporal modulation with spatial confinement.
CPA is achieved by coherently matching a single light source to a selected decaying Floquet mode.
By controlling the modulation frequency and amplitude, we identify the coalescence of two CPA branches at second-order EPs.
In this regime, the residual intensity of the growing Floquet mode exhibits a quartic dependence on momentum detuning, yielding a flattened absorption lineshape and enhanced robustness, as in conventional CPA EPs \cite{10.1103/PhysRevLett.122.093901, 10.1126/science.abj1028}.
Phase control further brings three CPA branches together at the wavevector of maximum Floquet rate, forming a third-order EP for which the topology provides a basis for coherent control of temporal CPA.
By transferring the source energy to the external modulation, temporal CPA EPs can operate below the material absorption edge, allowing for high energy efficiency and suppressed local heating.
These results indicate the great potential of photonic time crystals for coherent control of light in integrated nanophotonic devices.

% \section{Result}

\begin{figure}
    \includegraphics[width=\linewidth]{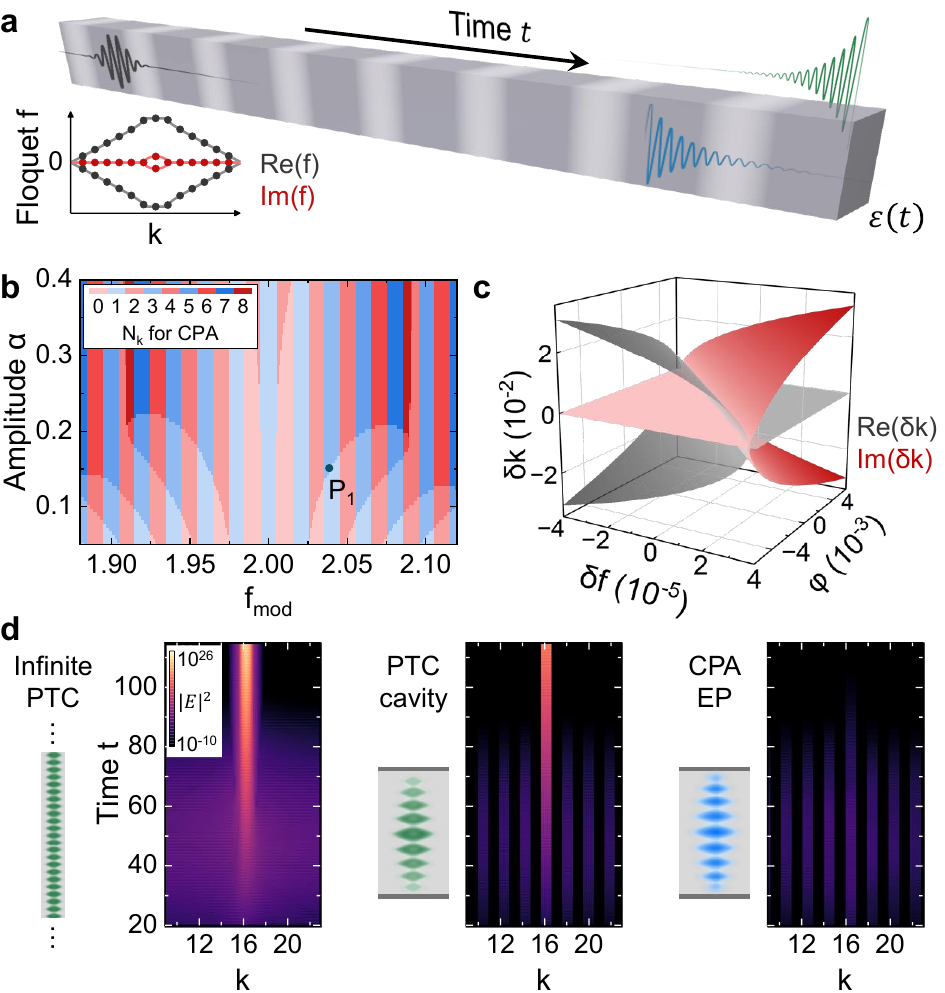}
    \caption{\label{f1}
        (a) Schematic of a photonic time crystal (PTC) with a periodically modulated permittivity $\varepsilon\left( t\right)$.
        A Gaussian pulsed light source is at the center. 
        The inset shows momentum bands (lines) for the infinite PTC and discrete wavevectors (dots) when the PTC is confined in a cavity.
        (b) Number $N_k$ of distinct real CPA wavevectors within the first momentum bandgap versus modulation frequency $f_{\mathrm{mod}}$ and amplitude $\alpha$ at $\varphi=0$.
        $N_k$ changes by 2 across a second-order temporal CPA EP.
        (c) Real and imaginary parts of the CPA wavevector detuning near P1 denoted in (b).
        (d) Evolution of momentum spectra from FDTD simulations for an infinite PTC (left) and a confined PTC in the non-CPA (middle) and CPA EP (right) cases.
    }
\end{figure}

We begin by discussing a one-dimensional photonic time crystal with the periodically modulated permittivity $\varepsilon(t)=\varepsilon_m\left[1+\alpha\sin(2\pi f_{\mathrm{mod}}t+\varphi)\right]$, as schematically shown by light and dark gray colors in Fig. \ref{f1}(a).
$\varepsilon_m=4$ is the average permittivity, $\alpha$ is the relative modulation amplitude, $f_{\mathrm{mod}}$ is the modulation frequency, and $\varphi$ is the modulation phase.
In a photonic time crystal with infinite length, the Floquet components at $f$ and $f-f_{\mathrm{mod}}$ couple resonantly near $f=f_{\mathrm{mod}}/2$, opening a momentum bandgap as shown by the lines in the inset of Fig. \ref{f1}(a) \cite{10.1364/AOP.382052, 10.1117/1.AP.4.1.014002}.
The positive (negative) imaginary part of the Floquet frequency corresponds to an exponentially growing (decaying) mode \cite{10.1126/science.abo3324, 10.1126/sciadv.abo6220, 10.1126/sciadv.adg7541}.
In contrast, when the photonic time crystal is confined in a cavity of length $l$, the spatial confinement gives rise to discrete momentum modes at $k_n=n\pi/l$ ($n=1,2,3,\ldots$), as shown by the dots in the inset.
This allows selective coupling of the light source to one momentum mode within the bandgap through control of the cavity length.

To establish the temporal CPA, a single light source is introduced at the center of the cavity.
The Gaussian source is $j(t)=\exp\left[-{(t-t_0)^2}/\left({2\tau^2}\right)\right]\cos\left(2\pi f_0 t\right)$ for $0\leq t\leq t_{\mathrm{e}}$ while $j(t)=0$ otherwise.
We use normalized units with $c=f_0=1$ and set $t_0=50$, $\tau=10$, and $t_{\mathrm e}=100$.
The growing and decaying Floquet modes of the source-free system are denoted by $\boldsymbol{g}(k,t)=(H_g,D_g)^{\mathsf T}$ and $\boldsymbol{d}(k,t)=(H_d,D_d)^{\mathsf T}$, respectively, where $H$ is the magnetic field and $D$ the electric displacement.
The optical response to a light source of unit amplitude is described by $\boldsymbol{v}_{\mathrm{src}}(k,t_m)=(H_{\mathrm{src}},D_{\mathrm{src}})^{\mathsf T}$, obtained from zero initial fields and evaluated at a time $t_m\geq t_{\mathrm e}$.
Temporal CPA requires a perfect match between the field generated by the source and the decaying Floquet mode at $t_m$, giving
\begin{equation}
    B(k;\boldsymbol{\lambda})
    \begin{pmatrix}
        u\\ c_d
    \end{pmatrix}
    =0,\qquad
    B=
    \begin{pmatrix}
        H_{\mathrm{src}} & -H_d\\
        D_{\mathrm{src}} & -D_d
    \end{pmatrix},
    \label{eq:f1_matching}
\end{equation}
where $\boldsymbol{\lambda}=(\alpha,f_{\mathrm{mod}},\varphi)$ denotes the modulation parameters, $u$ is the pulse amplitude, and $c_d$ is the amplitude of the matched decaying mode.
Matching both field components ensures that the field subsequently follows the decaying mode, so the temporal CPA condition is
\begin{equation}
    \mathcal{F}(k;\boldsymbol{\lambda})
    \equiv
    \det B(k;\boldsymbol{\lambda})=0.
    \label{eq:f1_cpa}
\end{equation}
After the source pulse ends, the optical response can be expressed as a superposition $\boldsymbol{v}_{\mathrm{src}}=F\boldsymbol{g}+C_d\boldsymbol{d}$, where $F$ and $C_d$ denote the excitation amplitudes of the growing and decaying Floquet modes, respectively.
For CPA, contributions to the growing mode excited at different times during the pulse interfere destructively, yielding $F=0$.
This is equivalent to the matching condition $\mathcal F=0$, as $\mathcal F=-F\det[\boldsymbol g,\boldsymbol d]$ and the two Floquet modes are linearly independent inside the gap.

In Fig. \ref{f1}(b), we present the number $N_k$ of distinct real CPA wavevectors within the first momentum bandgap as a function of $f_{\mathrm{mod}}$ and $\alpha$ at $\varphi=0$.
As the modulation parameters vary, two CPA branches can coalesce at a real wavevector, after which the corresponding wavevectors become complex, resulting in a net decrease of 2 in $N_k$.
Here, the CPA states coalesce, while the growing and decaying Floquet modes remain distinct.
Such a second-order EP of CPA satisfies
\begin{equation}
    \mathcal{F}=0,\qquad
    \partial_k\mathcal{F}=0,
    % \partial_k^2\mathcal{F}\neq0,
    \label{eq:f1_ep}
\end{equation}
with all $k$ derivatives taken at fixed $\boldsymbol{\lambda}$.
For example, point P1 in Fig. \ref{f1}(b) denotes a second-order CPA EP at $f_{\mathrm{P1}}=2.03934$, $\alpha_{\mathrm{P1}}=0.15$, and $\varphi_{\mathrm{P1}}=0$, with the corresponding wavevector $k_{\mathrm{P1}}=12.8439$.
$N_k$ changes from 3 to 1 across this EP.
In Fig. \ref{f1}(c), we present the coalescence of two CPA branches near P1 as a distinct feature of second-order EPs, where $\delta f=f_{\mathrm{mod}}-f_{\mathrm{P1}}$ and $\delta k=k-k_{\mathrm{P1}}$.

We use finite-difference time-domain (FDTD) simulations to verify the predicted temporal CPA EP.
Typical results are presented in Fig. \ref{f1}(d), and simulation details are provided in the Supplemental Material.
In the left panel of Fig. \ref{f1}(d), we present the evolution of momentum spectra in the photonic time crystal with infinite length, using $f_{\mathrm{mod}}=2.02$, $\alpha=0.15$, and $\varphi=0$.
In this case, the single light source excites a continuous momentum spectrum containing growing Floquet components within the bandgap,  consistent with the amplification reported in previous studies \cite{10.1117/1.AP.4.1.014002}.
The middle panel shows the results for a photonic time crystal confined in a cavity with a length $l=4.22391$ and perfect electric conductor (PEC) boundaries, while retaining the same modulation parameters.
As shown, spatial confinement gives rise to discrete cavity modes, and only the $n=17$ mode exists in the first momentum bandgap.
% The temporal modulation ($\boldsymbol{\lambda}$) and the spatial confinement ($l$) jointly determine whether that mode satisfies the CPA condition.
The selected cavity mode in the middle panel does not satisfy the CPA condition and therefore exhibits exponential growth, similar to that in the infinite photonic time crystal.
In contrast, in the right panel of Fig. \ref{f1}(d), we present the results with $l=4.15771$ and $f_{\mathrm{mod}}=2.03934$, while $\alpha=0.15$ and $\varphi=0$ are retained.
In this case, the wavevector of the selected $n=17$ mode matches that of the second-order CPA EP at P1 in Fig. \ref{f1}(b).
The field intensity decreases to approximately $10^{-8}$ at $t=100$ compared to its peak value, consistent with the predicted temporal CPA response dominated by the decaying mode.

\begin{figure}
    \includegraphics[width=\linewidth]{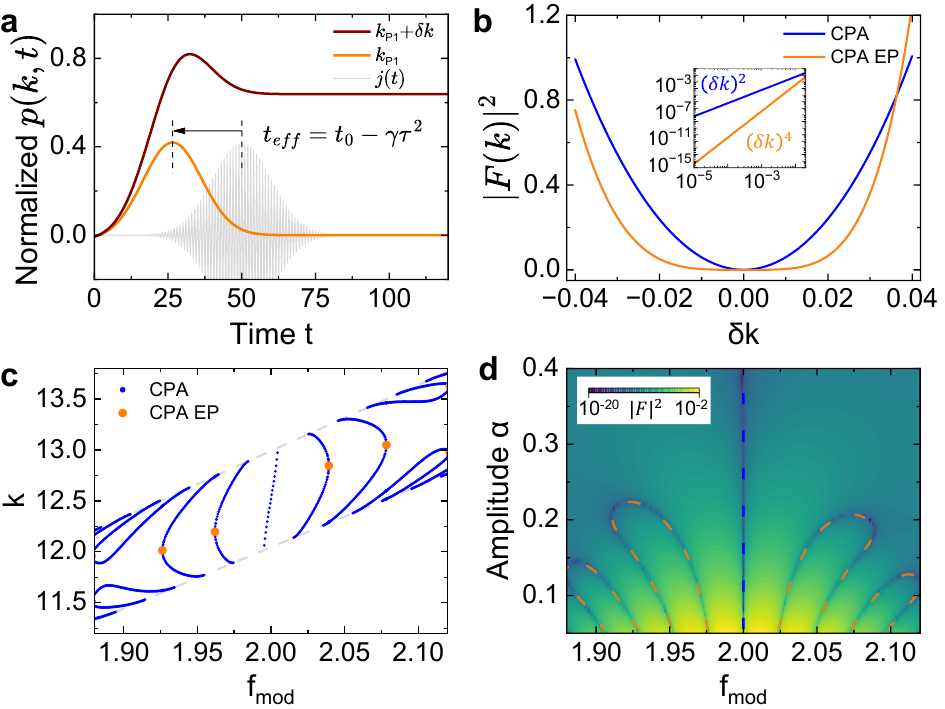}
    \caption{\label{f2}
        (a) Normalized cumulative excitation for the CPA EP at $k_{\mathrm{P1}}$ (orange) and the non-CPA case with a detuning $\delta k=0.02$ (brown).
        The gray line schematically shows the source $j(t)$.
        (b) Residual excitation intensity versus momentum detuning $\delta k$.
        The logarithmic inset shows quadratic dependence at CPA and quartic dependence at CPA EP.
        (c) Real CPA wavevectors in the first momentum bandgap.
        Blue dots correspond to CPA, and orange dots denote the CPA EPs.
        Gray dashed lines denote the band edges.
        (d) Residual excitation intensity at the bandgap center $k_*$, where the Floquet rate is maximal.
        The blue dashed line corresponds to CPA.
        Orange dashed lines denote CPA EPs in Fig. \ref{f1}(b).
    }
\end{figure}

To quantify the deviation from CPA, we introduce the cumulative excitation amplitude $p(k,t)$ of the growing Floquet mode, accounting for source contributions up to time $t$.
The final excitation amplitude is then given by $F(k)\equiv p(k,t_{\mathrm e})$.
We calculate $p(k,t)$ using the Green's function method (details in Supplemental Material) and compare CPA EP and non-CPA responses in Fig. \ref{f2}(a).
The orange, brown, and gray lines show the CPA EP response, non-CPA response, and source profile $j(t)$, respectively.
As shown, the exponential weighting of the source contributions shifts the effective Gaussian envelope toward earlier times, giving an effective center $t_{\mathrm{eff}}=t_0-\gamma\tau^2$, where $\gamma$ is the Floquet growth rate.
In both cases, $p(k,t)$ initially increases as the source excites the growing Floquet mode.
At the CPA EP, however, contributions to the growing mode from different times within the single pulse interfere destructively, finally bringing $p(k,t)$ back to zero.
In contrast, $p(k,t)$ remains nonzero after the pulse in the non-CPA case.
This destructive self-interference enables the field generated by a single pulse to match the decaying Floquet mode and establishes temporal CPA.

We demonstrate the enhanced robustness of a CPA EP against momentum detuning $\delta k$ by comparing the residual excitation intensity $\lvert F(k)\rvert^2$ with that of CPA ($\partial_k\mathcal{F}\neq0$).
Typical results are presented in Fig. \ref{f2}(b), where the orange and blue lines correspond to the CPA EP (P1) and CPA ($\alpha=0.15$, $f_\mathrm{mod}=2$, $k=12.5174$), respectively.
For small momentum detunings,  $\lvert F(k)\rvert^2$ is proportional to $(\delta k)^2$ for CPA and $(\delta k)^4$ for the CPA EP.
The quartic response at the CPA EP results in a flattened absorption lineshape as in conventional CPA EPs \cite{10.1103/PhysRevLett.122.093901, 10.1126/science.abj1028}.
This enhanced robustness against momentum detuning reduces the sensitivity of temporal CPA to small variations in cavity length.

Besides the enhanced robustness, CPA EPs near the bandgap center exhibit relatively fast absorption, owing to the large Floquet rate in this region.
In Fig. \ref{f2}(c), we present the CPA wavevectors within the first momentum bandgap as a function of modulation frequency $f_{\mathrm{mod}}$ at $\alpha=0.15$ and $\varphi=0$.
The blue and orange dots correspond to CPA and CPA EPs, respectively.
These CPA EPs are near the bandgap center $k_*$, where the Floquet rate is maximal.
We further explore absorption at the maximum rate by tuning the cavity length to align the selected mode with $k_*$.
In Fig. \ref{f2}(d), we present the residual excitation intensity $\vert F(k_*;\boldsymbol{\lambda}) \vert ^2$ obtained from FDTD simulations as a function of modulation frequency $f_{\mathrm{mod}}$ and amplitude $\alpha$ at $\varphi=0$.
The dark regions indicate strongly suppressed excitation of the growing mode.
The blue dashed line near $f_{\mathrm{mod}}=2$ corresponds to CPA at $k_*$.
The orange dashed lines denote CPA EPs identified in Fig. \ref{f1}(b) and agree well with the regions of strongly suppressed excitation at $k_*$.
This is because the robustness of nearby CPA EPs against momentum detuning allows the residual excitation at $k_*$ to remain strongly suppressed.

\begin{figure}
    \includegraphics[width=\linewidth]{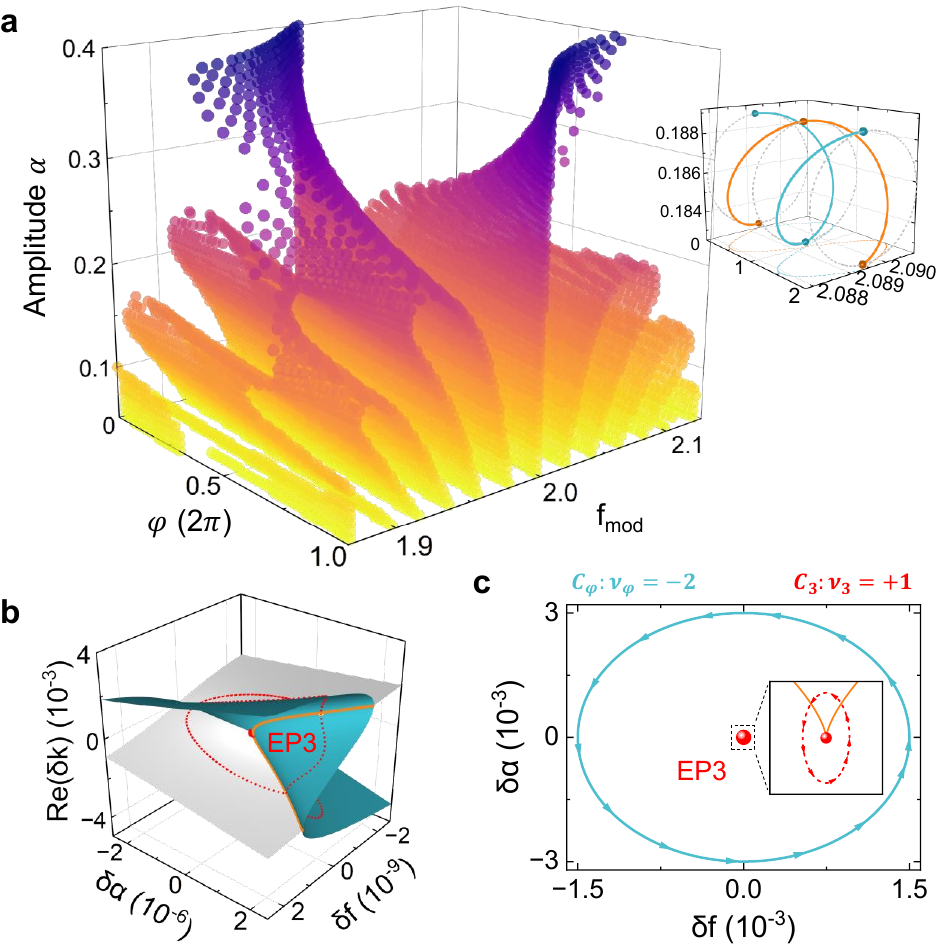}
    \caption{\label{f3}
        (a) CPA manifold at the maximum Floquet rate in the modulation parameter space $(f_{\mathrm{mod}},\alpha,\varphi)$.
        The inset shows two CPA trajectories around the third-order CPA EP (EP3).
        (b) Real parts of the relative CPA wavevectors near EP3 at $\varphi=\varphi_{\mathrm{EP3}}$.
        All coordinates are relative to EP3.
        The blue and gray sheets correspond to real and complex CPA wavevectors, respectively.
        The solid orange lines denote second-order CPA EPs, and the red dotted trajectories show the evolution along the parameter loop $C_3$.
        (c) Parameter loops around the EP3 in the $(\delta f,\delta\alpha)$ plane.
        The blue loop $C_\varphi$ is the projection of the blue CPA trajectory in the inset of (a).
        The inset shows the projections of trajectories in (b).
    }
\end{figure}

Moreover, we reveal a topological correlation between CPA at the maximum Floquet rate and a third-order CPA EP through phase control.
We extract the CPA manifold at $k_*$ from the condition $F(k_*;\boldsymbol{\lambda})=0$ in the three-dimensional modulation parameter space $(\alpha, f_{\mathrm{mod}}, \varphi)$ and present the results in Fig. \ref{f3}(a).
As $\varphi$ varies, the CPA branches switch between Riemann sheets, reflecting the nontrivial topology of the manifold \cite{10.1126/science.aar7709, 10.1103/PhysRevLett.86.787, 10.1038/nature18605}.
The inset of Fig. \ref{f3}(a) shows two trajectories of CPA at $k_*$ that exchange branches after $\varphi$ changes by $2\pi$ and return to their initial positions after $4\pi$.
The double phase winding is characterized by $\nu_\varphi=(2\pi)^{-1}\oint_{C_\varphi}\mathrm{d}\varphi_{\mathrm{CPA}}=-2$, where $\varphi_{\mathrm{CPA}}$ is continuously unwrapped along the loop $C_\varphi$.
We further identify a third-order CPA EP on this manifold at $\alpha_\mathrm{EP3}=0.185727$, $f_\mathrm{EP3}=2.08940$, and $\varphi_\mathrm{EP3}=0.0262864$, where three CPA branches coalesce at the wavevector $k_*$.
In addition to $F=\partial_kF=0$, this third-order CPA EP satisfies $\partial_k^2F=0$.
In Fig. \ref{f3}(b), we present the real parts of the relative wavevectors $\delta k=k-k_{\mathrm{EP3}}$ for the three CPA branches near this EP.
At the third-order CPA EP, the residual excitation intensity is $\lvert F\rvert^2 \propto (\delta k)^6$, resulting in a flatter absorption lineshape and better robustness than at a second-order EP.
The red dotted trajectories in Fig. \ref{f3}(b) show the evolution of the CPA wavevectors along the parameter loop $C_3$, which encloses the EP3 with a topological charge of $\nu_3=+1$ \cite{10.1103/PhysRevLett.127.186602, 10.1038/s41467-023-42414-z}.
The orange solid lines denote the second-order CPA EP branches that coalesce at this third-order EP.

This correlation originates from the phase compensation required to suppress the excitation of the growing Floquet mode.
The residual excitation can be expanded as $F(k_*;\boldsymbol{\lambda})
=\sum_{m=1,3,5,\ldots} \mathrm{Re}\left({C_m e^{i m\varphi/2}}\right)$, where $C_m$ is the complex excitation coefficient associated with the $m$th Floquet harmonic.
In the leading harmonic approximation, $F(k_*;\boldsymbol{\lambda})
\approx \mathrm{Re}\left({C_1 e^{i \varphi/2}}\right)$.
CPA then requires $\varphi_{\mathrm{CPA}}=\pi-2\arg C_1$.
This phase relation yields $\nu_\varphi=-2w(C_1)$, where $w(C_1)=(2\pi)^{-1}\oint_{C_\varphi}\mathrm{d}\arg C_1$ is the phase winding number of $C_1$.
In this scheme, $C_1$ vanishes at EP3, and its phase winding reproduces the local spectral winding, yielding $w(C_1)=\nu_3$ (details in Supplemental Material).
Therefore, $\nu_\varphi=-2\nu_3$, consistent with the winding numbers shown in Fig. \ref{f3}(c).
The third-order EP topology thus provides a basis for coherent control of temporal CPA.

\begin{figure}
    \includegraphics[width=\linewidth]{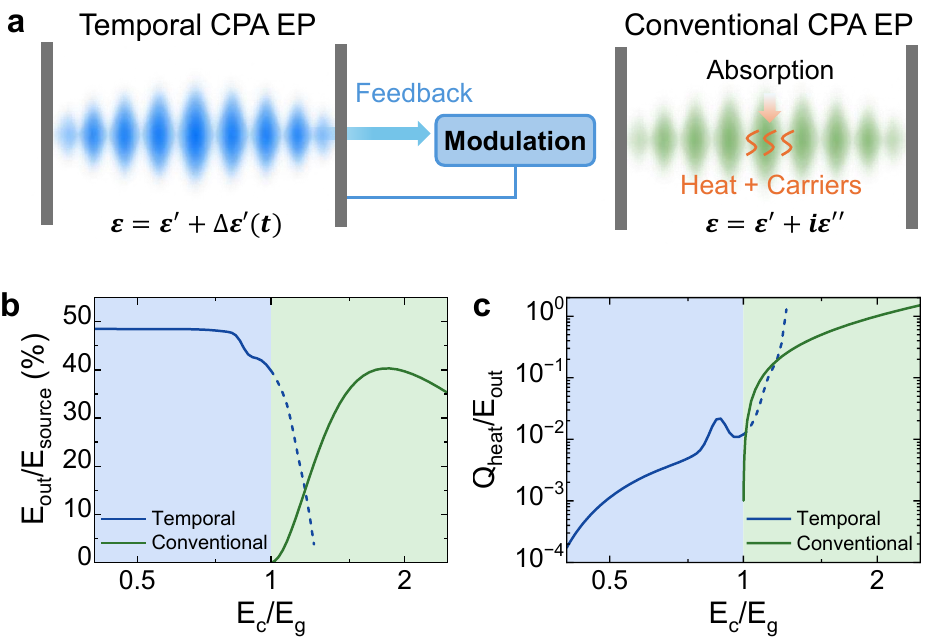}
    \caption{\label{f4}
        (a) Schematic of the temporal and conventional CPA EPs.
        The temporal CPA EP relies on temporal modulation of the real permittivity $\varepsilon=\varepsilon'+\Delta\varepsilon'(t)$.
        The conventional CPA EP relies on interband absorption associated with the imaginary part of the permittivity  $\varepsilon=\varepsilon'+i\varepsilon''$.
        (b) Energy transfer efficiency and (c) relative local heating for the two types of CPA EPs.
        The blue and green backgrounds in (b) and (c) indicate photon energies below and above the bandgap, respectively.
    }
\end{figure}

Temporal CPA EPs also provide a basis for high energy efficiency and suppressed local heating.
As shown in the left panel of Fig. \ref{f4}(a), temporal CPA EPs rely on the temporal modulation of the real part of the permittivity.
The source energy is fed back to the external modulation without local heating in an ideal lossless dielectric.
In contrast, the conventional CPA EP shown in the right panel relies on material absorption associated with the imaginary part of the permittivity.
For interband absorption, the photon energy in excess of the electronic bandgap is dissipated as local heat through carrier thermalization.

To compare the two CPA EPs, we use a Tauc–Lorentz dielectric with an electronic bandgap $E_g$ \cite{10.1063/1.118064} and a Gaussian pulse with a central photon energy $E_c$.
Within their respective operating ranges, both systems are tuned to maintain the CPA EP condition as $E_c/E_g$ varies (details in Supplemental Material).
For the conventional CPA EP, the incident energy $E_{\mathrm{source}}$ is partitioned into reflected energy $E_{\mathrm{back}}$, extractable carrier energy $E_{\mathrm{carrier}}$, and local heat $Q_{\mathrm{heat}}$.
For the temporal CPA EP, $E_{\mathrm{source}}$ denotes the energy supplied by the source, $E_{\mathrm{back}}$ the energy returned to it, and $E_{\mathrm{feed}}$ the net energy fed back to the external modulation.
Interband absorption also generates carriers and local heat when the photon energy or a Floquet sideband energy exceeds the bandgap $E_g$.
In Fig. \ref{f4}(b), we present the energy transfer efficiency $E_{\mathrm{out}}/E_{\mathrm{source}}$, where $E_{\mathrm{out}}=E_{\mathrm{carrier}}$ for the conventional CPA EP and $E_{\mathrm{out}}=E_{\mathrm{feed}}+E_{\mathrm{carrier}}$ for the temporal CPA EP.
Even at a CPA EP, reflected or returned energy and local heating can limit the energy transfer efficiency of a finite pulse.
The temporal CPA EP can achieve an efficiency of $48\%$ below the bandgap ($E_c/E_g<1$), as shown by the blue solid line.
The blue dashed line shows that the temporal CPA EP persists above the bandgap ($E_c/E_g>1$), although stronger interband absorption reduces the energy transfer efficiency.
The conventional CPA EP operates above the bandgap ($E_c/E_g>1$).
The efficiency reaches a maximum of $40\%$ at $E_c/E_g=1.8$, as higher photon energies increase local heating, whereas lower photon energies increase reflection.
By comparison, the temporal CPA EP can achieve an energy transfer efficiency approximately $20\%$ higher than the maximum efficiency of the conventional CPA EP.
In Fig. \ref{f4}(c), we present the relative local heating $Q_{\mathrm{heat}}/E_{\mathrm{out}}$.
Owing to suppressed interband absorption, the local heating in the temporal CPA EP can be nearly three orders of magnitude lower than that in the conventional CPA EP.

% \section{Summary}

In summary, we demonstrate temporal CPA EPs in the momentum bandgap of a spatially confined photonic time crystal.
These CPA EPs exhibit flattened absorption lineshapes and enhanced robustness, while the proximity to the bandgap center enables fast absorption.
The third-order EP topology further provides a basis for coherent control of temporal CPA and for exploring EP dynamics.
Meanwhile, energy feedback to the external modulation can lead to higher energy transfer efficiency and suppressed local heating.
These results open new avenues for non-Hermitian Floquet photonics and indicate potential applications in integrated nanophotonic devices.

\begin{acknowledgments}
    This work is supported by the National Natural Science Foundation of China (Grants No. 12494600, 12494601, 12494603, and 12474426) and the Chinese Academy of Sciences Project for Young Scientists in Basic Research (Grant No. YSBR-112).
\end{acknowledgments}

% \clearpage

\bibliography{refer}

\end{document}